\documentclass{emulateapj}

\usepackage{graphicx}
\usepackage{epstopdf}

\newcommand{\lx}{erg~s$^{-1}$}

\newcommand{\chandra}{{\it Chandra}}
\newcommand{\xmm}{{\it XMM-Newton}}
\newcommand{\rxte}{{\it RXTE}}

\newcommand{\ppuls}{$P_{pulse}$~}
\newcommand{\porb}{$P_{orbit}$~}

\shorttitle{Capture-Recapture}
\shortauthors{Laycock}

\begin{document}


\title{From Blackbirds to Black Holes: Investigating Capture-Recapture Methods for Time Domain Astronomy}


\author{Silas G. T. Laycock \altaffilmark{1}}
\affil{Lowell Center for Space Science and Technology, 600 Suffolk St, Lowell, MA, 01854}

\altaffiltext{1}{Department of Physics and Applied Physics, Olney Science Center, University of Massachusetts Lowell, MA, 01854, USA}



\begin{abstract} 

In time domain astronomy, recurrent transients present a special problem: how to infer total populations from limited observations. Monitoring observations may give a biassed view of the underlying population due to limitations on observing time, visibility and instrumental sensitivity.   A similar problem exists in the life sciences, where animal populations (such as migratory birds) or disease prevalence, must be estimated from sparse and incomplete data. The class of methods termed Capture-Recapture is used to reconstruct population estimates from time-series records of encounters with the study population. This paper investigates the performance of Capture-Recapture methods in astronomy via a series of numerical simulations. The {\it Blackbirds} code  simulates monitoring of populations of transients, in this case accreting binary stars (neutron star or black hole accreting from a stellar companion) under a range of observing strategies. We first generate realistic light-curves for populations of binaries with contrasting orbital period distributions. These models are then randomly sampled at observing cadences typical of existing and planned monitoring surveys.  The classical capture-recapture methods, Lincoln-Peterson, Schnabel estimators, related techniques, and newer methods implemented in the {\it Rcapture} package are compared. A general exponential model based on the radioactive decay law is introduced, and demonstrated to recover (at 95\% confidence) the underlying population abundance and duty cycle, in a fraction of the observing visits (10-50\%) required to discover all the sources in the simulation. Capture-Recapture is a promising addition to the toolbox of time domain astronomy, and methods implemented in {\it R} by the biostats community can be readily called from within {\it Python}. 
\end{abstract}

\section{Introduction}
Thanks to the decade-plus durations of high energy missions such as \chandra, \xmm, {\it Integral}, and \rxte, long-term X-ray monitoring campaigns have been conducted for a number of galaxies and star clusters. In the Milky Way, most of this monitoring effort has been focussed on the Galactic Center (e.g. \citealt{markwardt2010}, \citealt{muno2004}, \citealt{hong2011}, \citealt{wijands2006}, \citealt{kulkers2007}); Galactic Plane (e.g. \citealt{grindlay2005}, \citealt{motch2010}); and the Orion Nebula \citep{coup}. Beyond the galaxy, extended monitoring campaigns have been performed for the Small Magellanic Cloud (e.g. \citealt{galache2008}, \citealt{eger2008}, \citealt{coe2010}) and the local group galaxies M31 \citep{williams2006}, M33 \citep{williams2008}, and M81\citep{sell2011}.  Such surveys  turn up large numbers of transient sources belonging to astrophysically interesting classes, such as flaring stellar coronae, and accreting binaries containing black holes, neutron stars and white dwarfs. The present work is motivated by a natural question posed by this profusion of new transients: ``How to count a population of recurrent transient sources?"

Due to their underlying physics, high energy sources tend to exhibit very large changes in luminosity on virtually all timescales. Although no set definition of ``transient" exists, the term is universally applied to objects that appear in only a subset of repeat observations (of the same sky region), and are absent in others, despite sufficient sensitivity to detect them. The types of variability that result in the classification of ``transient" are characterized by large amplitude (factor of 10 or more), and prolonged periods of quiescence (days-years) between high luminosity states.  Accreting Binaries (see e.g. \citealt{xraybook}) and Flare Stars \citep{favata2003} form the two largest classes of recurrent transients in the X-ray regime. High mass X-ray binaries (HMXBs) show large-amplitude outbursts (duration: days-weeks) at intervals related to the orbital period (months). The strength of these outbursts is frequently modulated by slower (years) changes in the mass outflow rate of the companion star. The exact interplay between these timescales is tied to complex physical interactions between the two stars  which are beyond the scope of this paper (see e.g.  \citealt{reig2011}) for a review). Suffice to say that HMXB activity can be modulated in a quasi-periodic way, or completely suppressed.  Low mass X-ray binaries (LMXBs) and Cataclysmic Variables (CVs) tend to remain in a persistent low luminosity or quiescent state, punctuated by rare outbursts of weeks-months duration. These events: bursts, X-ray novae, dwarf novae, classical novae, are not periodic, but are thought to recur on certain characteristic timescales \citep{mukai2008}. Active galactic nuclei (AGN) can also be considered transients with very long timescales (decades - millennia). At optical wavelengths a zoo of irregular variables, pulsating stars, eclipsing binaries etc. might appear transient under certain observing conditions (low cadence, low sensitivity). Supernovae and Gamma Ray bursts do not count, being strictly one-time events.

Finding the total population of any of the above classes of transients in a cluster or galaxy, is thus an unsolved problem, current cases of interest include the SMC, in which $\sim$100 HMXBs are currently known, and the overabundance of X-ray transients in the Galactic Center \citep{muno2005}. 

The astronomical transient-counting problem remains relatively unexplored, despite rigorous mathematical analysis of long-duration surveys dating back to at least \cite{Herschel1783}. Existing long-duration datasets on the whole sky include the Harvard photographic plate collection (1880s-1980s) currently undergoing digitization and data-mining by the DASCH\footnote{http://dasch.rc.fas.harvard.edu} collaboration \citep{Grindlay2012,Laycock2010}; current facilties generating large amounts of such observations include  ASAS/ASSAS-SN ~\citep{Pojmanski2002,Holoien2017}, Palomar Transient Factory~\footnote{http://www.ptf.caltech.edu}, MASTER ~\citep{Lipunov2010}, and the upcoming Large Synoptic Survey Telescope (LSST)\footnote{https://www.lsst.org}.    

The problem of deducing population parameters from repeated isolated observations is not unique to astronomy, and in fact an established approach to similar problems has been long-used in field-biology: Capture-Recapture analysis. In recent years, sophisticated variants of this method have been developed in the fields of ecology, epidemiology and biostatistics. Typical applications involve: small animal live-trapping (hence the term capture-and-recapture); estimation of fish populations from catch data \cite{schnabel1938}; migratory bird survival rates (e.g. Blackbirds. \citealt{miller2003}) from tagging studies; and disease prevalence inferred by comparing patient IDs in the medical records of multiple clinics. The shared feature of all these situations is an unknown population of identifiable individuals, who are encountered or observed on multiple randomly spaced occasions. Each encounter will yield a ratio of known(old) to unknown(new) individuals. Modeling of this ratio and its time evolution is the basis of capture-recapture methods. An introduction to many of these methods can be found in \cite{Amstrup2005}.

The aim of this paper is to explore the application of capture-recapture methods to astronomical transients. Section~\ref{sect:equations} outlines the statistical methods, then in section~\ref{sect:pulsarmodel} a numerical simulation code ({\it Blackbirds}) for populations of astronomical transients is described. The results of applying capture-recapture methods to recover the input model parameters, for a wide range of source properties and observing cadences, are presented in section~\ref{sect:results}.

\section{Review of Capture-Recapture Methodology and Algorithms}
\label{sect:equations}

\subsection{Lincoln-Petersen Index}
\label{sect:LP}
If a fixed population of individuals (which can be anything) is observed  on multiple occasions,  at each epoch the sample tends to contain a mixture of individuals who have been previously encountered, and some who are new. In the simplest case, where all individuals have the same probability of being encountered at any time, the total population can be deduced from the ratio of ``new" to ``old" individuals encountered at two successive epochs. This mathematical relationship known as the Lincoln-Petersen index, is given by Equation~\ref{eqn:lp}, where $N_{LP}$ is the total population estimate derived from $N_1$ (individual encountered on occasion 1), $N_2$ (individuals encountered on occasion 2) and $N_{1,2}$ (individuals encountered on both occasions). The modified form due to \cite{chapman1951} avoids divergence in the case of small numbers, and is normally distributed (provided the assumptions of sample independence and capture probability are met). The modified Lincoln-Petersen index, and its 1$\sigma$ uncertainly are given by Equations~\ref{eqn:lpfull}~\&~\ref{eqn:lpvar}. 

\begin{equation}
N_{LP} \simeq \frac{N_1 \times N_2}{N_{1,2}}
\label{eqn:lp}
\end{equation}

\begin{equation}
N_{LP} = \frac{(N_1+1)  \times (N_2 + 1)} {(N_{1,2} + 1)} - 1
\label{eqn:lpfull}
\end{equation}

\begin{equation}
\sigma^2_{LP} = {\frac{(N_1+1)(N_2+1)(N_1-N_{1,2})(N_2-N_{1,2})}{(N_{1,2} + 1)(N_{1,2} + 1)(N_{1,2} + 2)}}
\label{eqn:lpvar}
\end{equation}

When more than two samples are available, Equation~\ref{eqn:lpfull} can be applied to adjacent pairs, and a running average constructed. This approach does not make maximum use of the available information, although it does tend to converge. Provided the conditions of closed population and equal capture probability apply across individuals and observations it will give an unbiassed estimate.  

\subsection{Schnabel  Estimator}

More efficient use of monitoring data is made by the Schnabel Estimator \citep{schnabel1938} which is essentially a cumulative weighted average of the ratio of new individuals to previously encountered individuals. 

\begin{equation}
Ns_i = \frac { \sum{N_i \times Nc_{(i-1)} }} { (\sum{Nr_i}) + 1 } 
\label{eqn:schnabel}
\end{equation}

Equation~\ref{eqn:schnabel} gives the running population estimate $Ns_i$ in terms of the 
number of individuals ($N_i$) encountered on occasion {\it i}, the cumulative number of individuals encountered ($Nc_(i-1)$) prior to occasion {\it i}, and the number of individuals re-encountered ($Nr_i$) on occasion {\it i}.   In programming this, we avoid finding $Nr$ explicitly since it can be shown that $Nr_i$ = $N_i$ + $Nc_{(i-1)}$ - $Nc_i$. A formula for the uncertainty in $Ns$ is given by \cite{zar1996}, however for small $Nr$ one should use the poisson distribution instead. In astronomy we are also accustomed to deriving confidence limits by Monte Carlo simulation.   

\subsection{Derivation of a Generalized Exponential Capture-Recapture Function}

While many {\it ad hoc} refinements to the above methods can be found in the literature, the birth of modern techniques lies in the realization that encounter history should be modeled using a probability function. Under the condition of ``equal capture probability", (all individuals have the same probability of being in an observable state at any given instant; provided the observations are widely spaced and uncorrelated) it follows that the rate of encounters with new (previously undetected) individuals will follow an exponential curve since the number of undetected individuals falls with successive observation; the situation being closely analogous to radioactive decay, with the encounter probability replacing the decay constant.

Assume a star cluster contains $N_0$ transients, all having the same probability $p$ of being ``On" during any observation.  
For each transient, $p$ can be identified as the star's duty cycle, since over sufficiently long time intervals, the ratio of time spent in outburst, to time in quiescence {\it is} the probability of the star being in outburst at some arbitrary time. This result applies to both periodic and aperiodic transients, and is independent of the actual timescales involved.

If this population is observed many times, then at each observation $k$  (where k=1,2,3,4......) we see a mixture of `new' and `old' stars. At any such observation, we can expect to see $n_k$=$N_0 p$ stars (with large uncertainty). Subsequent observations increase the cumulative number ($Nc_k$) of stars encountered, thus reducing the number of remaining {\it un-encountered}  stars ($Nu_k$). Following the radioactive decay law as a model leads to Equation~\ref{eqn:decay}.

\begin{equation}
\label{eqn:decay}
Nu_k = N_0 e^{-pk}
\end{equation}

The cumulative number of stars encountered by the time one reaches the $k$th observation will be: 

\begin{equation}
 Nc_k = N_0 - Nu_k
\end{equation}

Thus, 
\begin{equation}
Nc_k = N_0(1-e^{-pk})
\label{eqn:exponential}
\end{equation}

Hence a plot of cumulative number of transients $Nc_k$ vs observation number $k$ should be fit by Equation~\ref{eqn:exponential} to derive values for $N_0$ and $p$. This method is applicable to any number of observations ($k>3$). Some advantages over the LP and Schnabel estimators include better use of information, uncertainty estimates generated by the fitting process (e.g. least squares), and extensibility. Scientifically the greatest gain for astronomical purposes is direct estimation of the duty cycle.  

We note that in adopting Equation~\ref{eqn:exponential} we have implicitly treated $k$ as a continuous variable because Equation~\ref{eqn:decay} follows from a differential equation. An alternative derivation that does not include this feature is as follows.

Independently of the value of $k$, the total number of stars is given by $N_0 = N_{u_k} + N_{c_k}$ so the expected number of identified stars can be written as $n_k = p  N_0 =  p N_{u_k} + p N_{c_k}$. Now, before the k-th survey starts, the expected number of  unidentified stars is equal to $N_{u_{k-1}}$. When the k-th survey is completed,  the expected number of stars which are identified for the first time is given by  $p N_{u_{k-1}}$. \\

Hence, $N_{u_k} = N_{u_{k-1}} - p N_{u_{k-1}}$ or \\

$N_{u_k} = N_{u_{k-1}} (1 - p) $  \\

This is a ``difference equation" that, with the condition $N_{u_0} = N_0$, has solution given by  \\

$N_{u_k} = N_0 (1-p)^k$

From this last equation, one obtains  Equation~\ref{eqn:altform}, \\

\begin{equation}
\label{eqn:altform}
N_{c_k} = N_0 \left[1 - (1-p)^k \right]  \\
\end{equation}

which is functionally equivalent to Equation~\ref{eqn:exponential} for small values of $p$.  The exponential approximation converges more slowly resulting in an initial underestimation which diminishes with $k$. For $p$=0.1, the maximum discrepancy is less than 5\%.

\subsection{The Problem of Heterogeneous Capture Probability} 
Equation~\ref{eqn:altform} still does not make use of all the information  available, since it ignores the number of repeat detections of each source; making no distinction for example between recurrent, and one-time transients. Data can be used to furnish the form of such heterogeneity in the duty cycle $p$, with important astrophysical implications. If distinct sub-populations exist, each characterized by a unique $p_j$ then one can fit a multi-component model like Equation~\ref{eqn:multicomp} based on replication of Equation~\ref{eqn:altform} but if $p_j$ is a function of some other physical variable (age, orbital period, mass, magnetic field, etc) then a more careful analytic prescription will be needed.

\begin{equation}
\label{eqn:multicomp}
\{N_{c_k}\}_j = \sum_j N_{0,j} \left[1 - (1-p_j)^k \right]  \\
\end{equation}

The frequency of stellar flares turns out to be such a case. In a review paper \cite{Ambartsumian1975} summarizes their long-term study of the Pleiades star cluster (originally published in Russian in a series of 5 papers: \cite{Ambarts1970} -- \cite{Miro1977}).   They were able to show that the frequency of stellar flares ($f$) in young stars declines with age, although repeat photographs of the cluster are on a timescale much too short to show any individual star evolving. The total population of flare stars could be reproduced from the observed incidence of flares, {\it only}, if the rate of flaring varied between stars and correlated with an independent measure of stellar age. We reproduce their estimator here as Equation~\ref{eqn:ambart1} where $n_f$ is the number of flare stars for which $f$ flares have been observed, $\nu$ is the flare frequency and $t$ is the total effective observing time. Their quantity $\nu t$ is analogous to the quantity $pk$ in our notation, and is obtained directly from the data via a Lincoln-Petersen like identity $\nu t = 2n_2/ n_1$ using the numbers of stars ($n_1$, $n_2$) with one and two observed flares. In Equation~\ref{eqn:ambart2} there are $j$ sub-populations each with its own flaring frequency. 

\begin{equation}
\label{eqn:ambart1}
  n_f = N_0 e^{-\nu t} \frac{(\nu t)^f}{f!}  \\ 
\end{equation}

\begin{equation}
\label{eqn:ambart2}
  n_f = \sum_j N_{0,j} e^{-\nu_j t} \frac{(\nu_j t)^f}{f!}  \\ 
\end{equation}

A most useful analytic result obtained by ~\cite{Ambartsumian1975} is the following inequality describing the limits of $N_0$ in the case of heterogenous $\nu$. Although their parameterization of the problem differs slightly, it follows from the same axioms as the Capture-Recapture methods, and therefore Equation~\ref{eqn:ambart2} should be widely applicable. 

 \begin{equation}
\label{eqn:ambart2}
\frac{n_1^2}{2n_2} \leq N_0  \leq  \frac{n_1^2}{n_2} \\ 
\end{equation}

\subsection{Biostatistical Implementations: Rcapture}

Software for Capture-Recapture studies used in the bio-statistical field  is freely available, for example {\it Mark} \footnote{http://warnercnr.colostate.edu/~gwhite/mark/mark.htm}, {\it M-Surge} \citep{msurge}, {\it CARE} \citep{chao} and {\it RCAPTURE} \citep{rcapture}. These newer packages use capture histories from multiple epochs to iteratively build probabilistic models to describe the population and biases. The models are then used to obtain statistical bounds on the population, and to study the capture-probability distributions themselves, which can represent survival rates and other demographic parameters. The newer algorithms make use of landmark advances in statistical computing and are computationally intensive.  {\it Mark} dominates animal population studies and incorporates the widest range of algorithms, but most are inappropriate for astronomy. Similarly {\it M-Surge} is aimed at estimating survival and migration rates in open populations.  {\it CARE} is widely used in epidemiology. {\it Rcapture} is geared toward closed populations, including those with high degrees of heterogeneity. It incorporates methods from biology, medicine and census analysis in a generally applicable framework, and consequently was selected for use in this study.  {\it Rcapture} is conveniently distributed as a module for the widely used open source statistical programming platform {\em ``R"} \citep{R}.  

In the following sections we test Capture-Recapture methods from the basics up to the latest statistical approaches, against a suite of simulations designed to mimic astronomical monitoring campaigns.

\section{Simulating a Population of Transient X-ray Binaries: the {\it Blackbirds} code.}
\label{sect:pulsarmodel}
In order to provide trial datasets for analysis with capture-recapture methods, the {\it Blackbirds}  code was created to simulate populations of recurrent transients.
By using simulated data, the fidelity of the capture-recapture techniques be quantitatively assessed. The simulation enables one to model a wide range of possible source populations, each characterized by different recurrence timescales, and to control for the effects of sampling (observing cadence).  {\it Blackbirds} is written in {\it Fortran 90}, and is designed to generate a set of representative X-ray binary light-curves, `observed' via a simulated sampling structure. The type of source was chosen for its relevance to the author's observational work, and because the methods developed here will be applicable to existing monitoring campaigns.  Other types of recurrent transient can also be simulated by the code. 

\begin{figure}
\begin{center}
\includegraphics[width=5.5cm, angle=-90]{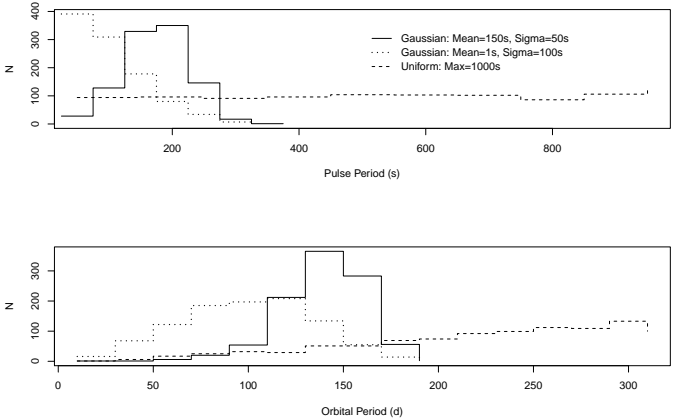}
\caption{Model Period Distributions. The upper panel shows 3 example \ppuls distributions, while the lower panel shows the resulting \porb distributions assuming the two are coupled by Equation~\ref{eqn:corbet}, as inferred from the Corbet diagram. }
\label{fig:period_distr}
\end{center}
\end{figure}

We could specify the distribution of recurrent time-scales directly, or via the distribution of another control parameter which is in principle better constrained by observation.  In the case of HMXB pulsars, the pulse period distribution is known somewhat independently of the orbital period-distribution (since \ppuls is obtained in even a single detection).  First, pulse periods are drawn at random from either: a uniform distribution; a gaussian distribution characterized by a mean period $\bar P$ and a width $\sigma$; or a gaussian truncated at its peak to produce an exponential. These model parameters are given in Table~\ref{tab:models} and the resulting distributions are plotted in Figure~\ref{fig:period_distr}, (upper panel). The gaussian parameters P1=150s, $\sigma$=100 were chosen to approximately reproduce the pulse period distribution observed by the {\it RXTE} monitoring project \cite{coe2010}. The other choices were selected to provide populations with contrasting recurrence timescales. Choice of distribution depends on the application, and can be specified to explore the observable consequences of various underlying population models.

\begin{figure*}
\begin{center}
\includegraphics[width=10cm, angle=-90]{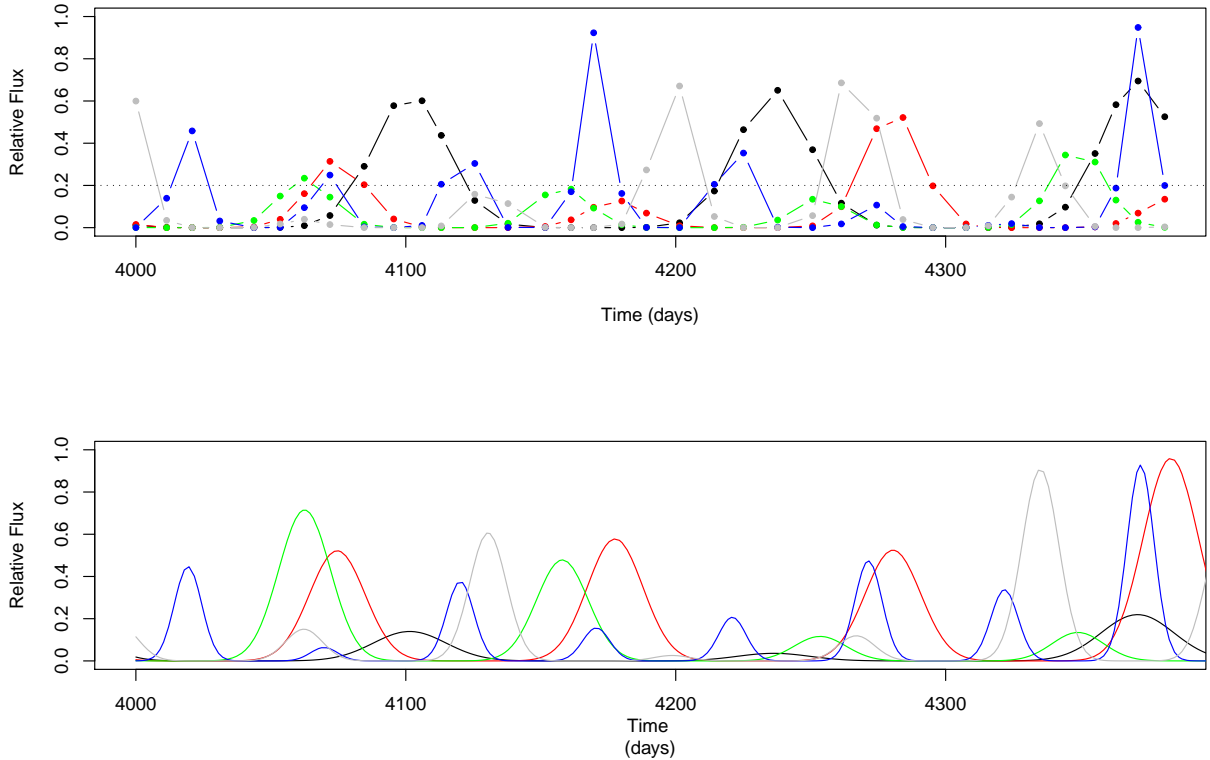}
\caption{Simulated Pulsar lightcurves. Lower panel shows 5 pulsars with a range of orbital periods, undergoing periodic oubtbursts characterized by a gaussian model with $\sigma$=0.1 in phase, and randomly  assigned peak flux. The upper panel shows a similar model population sampled at the nominal observing  cadence of the RXTE monitoring project: random intervals in the range 7-14 days. } 
\label{fig:model_lc}
\end{center}
\end{figure*}

Next, an orbital period is assigned to each X-ray binary using a fit to the Corbet diagram (\citep{corbet1984}) using up-to date values for 9 Galactic Be/X pulsars. The formula used is given in Equation~\ref{eqn:corbet}. Due to the logarithmic relationship between \ppuls and \porb, the orbital period distribution is a skewed gaussian for an input gaussian \ppuls distribution, and a monotonically rising function of \porb for a flat \ppuls distribution, as illustrated in Figure~\ref{fig:period_distr} (lower panel). A steep \ppuls distribution rising toward small values, modeled by a truncated gaussian with its mean value at or below the cut-off also produces a bell-curve in \porb.

\begin{equation}
log_{10}(P_{orbit}) = 1.1329 + 0.4532\times log_{10}(P_{pulse})
\label{eqn:corbet}
\end{equation}

Epoch of periastron for each individual system is randomly assigned within a 1000 day range preceding the beginning of the simulation. Outbursts of a given source occur at periastron, simulated by a gaussian profile centered at phase=0, with width ($\sigma$=0.1) selected to mimic the folded orbital profiles of \cite{galache2008}. The peak flux of each outburst is assigned to be a baseline value (1.0 in the simplest case) modified by a random deviate in the range 0-1.0.  The {\it RXTE} lightcurves in LO5/G08 show no relationship between pulse period and X-ray luminosity, hence the use of a simple uniform distribution for peak outburst flux. The imposition of this flux scaling factor is a key part of the simulation, as it injects a realistic aperiodic component into the long term lightcurve of each source.

\begin{table}
\caption{Pulsar Population Models.}
\begin{tabular}{lllll}
Model       &   Type    &     P1    &     P2      &   T     \\
		& 		&  s         &      s         &  days  \\		
\hline    
A               &  G          &     150  &     50       &    132  \\
 B              &  U          &     1       &     1000   &   $<$311      \\
 C             &  G          &     0        &     200     &    16  \\
D              &    G         &    200   &    100       &    150  \\
E               &   G         &     200   &   50           &   150  \\
F                &   G         &    0        &  100        &   16   \\
\hline
\end{tabular}
\medskip

Pulse period distribution Type is either Gaussian with $\bar{P_s}$=P1, $\sigma$=P2, or 
Uniform with $P_s $$<$P2. All models are truncated at $P_s$$>$1. The typical recurrence time $T$ is the corresponding most-common orbital period, derived from Equation~\ref{eqn:corbet}.
\label{tab:models}
\end{table}

Sampling intervals are assigned as a series of uniform-randomly distributed intervals with a specified minimum and maximum separation, or or gaussian-random with specified mean spacing and spread. This approach facilitates exploring the effects of varying the observing cadence. The cadences used in this work are: 7-14 days, 15-30d, 60-90d, 90-120d. The shortest interval is to comply with the {\it RXTE} monitoring project's ``weekly" observing cadence; in practice observations are unevenly spaced within 7-14 day intervals. 

Finally the model XRBs are sampled at the observing cadence (whether simulated or an actual series of tabulated times), modulo an imposed detection threshold, to produce simulated lightcurves for analysis. An example run, using 5 pulsars is shown in Figure~\ref{fig:model_lc}. Taking a `normal' (or type I) outburst to peak at $\sim$10$^{37}$\lx, and the sensitivity limit for RXTE to be $\sim$10$^{36}$\lx, a suitable simulation threshold is in the range 0.1 to 0.5 in the relative flux scale (0-1). 

The output of a single run of the simulation is a capture history. This is an array containing one column per source, and one row per observation, with each cell having a value of ``1" if the source is detected, and ``0" if not detected. This format follows the standard practice in biology.

Further refinements include the ability to specify different baseline flux normalizations to each source, and to parameterize the aperiodic component. These changes will be used to create model populations to be confronted with actual observational data. In particular: to simulate differential sensitivity, the fact that in real observations pulsars lie at different distances from the field center; To model the effects of period-flux correlation; and to explore the nature of long-term variability. A module to produce aperiodic source populations is also in development.

The simulation is not specific to X-ray pulsars (other than in the choice of pulse-period as the independent variable) , and can readily produce model datasets for any population of periodic sources. In typical use, populations of 50-100 sources are generated and analyzed (as described below), with 1000 such iterations being used to accumulate statistical bounds.

\section{Applying Capture-Recapture Methods to the XRB Simulation}
\label{sect:results}
The purpose of exploring capture-recapture techniques in this paper is to discover if robust constraints can be placed on the total number of transient or variable astronomical objects. For example pulsars in the SMC, based on the RXTE weekly monitoring observations. The motivation for this example is that after 10 years RXTE continued to discover new pulsars, and the rate of discovery must at some level be tied to the size of the underlying population. We begin with straightforward implementation of the Lincoln-Petersen index (Equation~\ref{eqn:lpfull}), explore its performance on our pulsar simulation under a range of sampling strategies and population distributions. We then move on to the Schnable estimator, which makes more complete use of information, and is formulated to incorporate multiple samples. Finally we examine sophisticated modern capture-recapture algorithms developed for use in wildlife ecology and epidemiology. Researchers in these fields have developed refinements to the basic concept that extend it to large numbers of visits (observations), individuals with unequal capture probabilities and correlations between samples.  The software package {\it Rcapture} \cite{rcapture} provides an implementation of many of these techniques, and provides example analyses of a wide variety of datasets drawn from biology and epidemiology. The poisson regression approach followed in {\it Rcapture} yields robust confidence intervals in addition to abundance estimates. 

\subsection{Implementation of the Lincoln-Petersen Estimator}
\label{sect:lpe}
\begin{figure}
\begin{center}
\includegraphics[width=5cm, angle=-90]{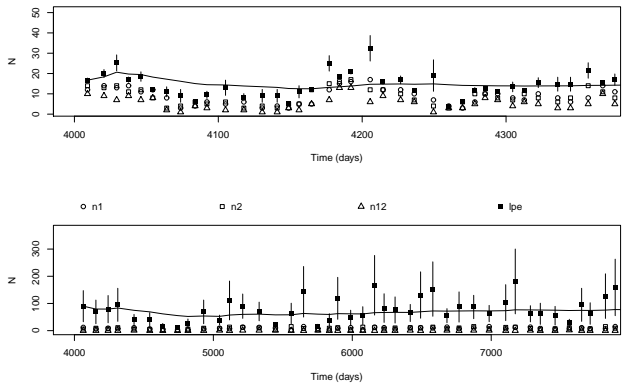}
\caption{Capture histories for source population model A, under two contrasting observing cadences. At each observation time, the code keeps track of the number of detected sources (n2),  the number detected in the previous observation (n1), and number of sources detected in both the current $and$ previous observation (n12). The Lincoln-Peterson estimator (Equation~\ref{eqn:lpfull}) and its running mean (solid line) are also plotted.
Observation times are drawn from a gaussian-random distribution of intervals. Mean sample interval $~$P (top), and $>>$P (bottom).  }
\label{fig:lpe_vs_time}
\end{center}
\end{figure}

We begin by computing the modified Lincoln-Petersen index (Equation~\ref{eqn:lpfull}) by treating our simulated lightcurves in a pairwise manner. For each consecutive pair of observations, we calculate $N_{LP}$, and also keep a cumulative running average. In this way we find out whether the estimator can converge on a stable value, and how rapidly. The result of such an approach is shown in Figure~\ref{fig:lpe_vs_time}, for a population of 1000 pulsars under two alternative period distributions, and a range of different sampling cadences.

It is evident from Figure~\ref{fig:lpe_vs_time} that such a simple implementation does a poor job in terms of consistency. Furthermore the result gets worse (not better) as the sampling density increases. The reason is that pairing consecutive samples violates the requirement that samples be independent, because the flux values are correlated on timescales comparable to the cadence. Simple application of Equation~\ref{eqn:lpfull} is therefore useful only when a small number of widely spaced observations are available.

\begin{figure}
\begin{center}
\includegraphics[width=5.3cm, angle=-90]{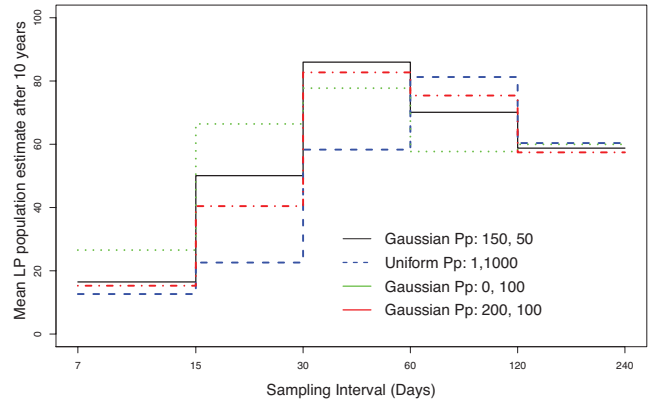}
\caption{Effect of Sampling Interval and Period Distribution on LP population estimator averaged over 10 years of monitoring data.  Four contrasting \ppuls models are compared, as in Table~\ref{tab:models}. Each bar in the histogram is the average of 10 independent simulation runs.}
\label{fig:lpe_sampling_model}
\end{center}
\end{figure}

\begin{figure}
\begin{center}
\includegraphics[width=9cm, angle=0]{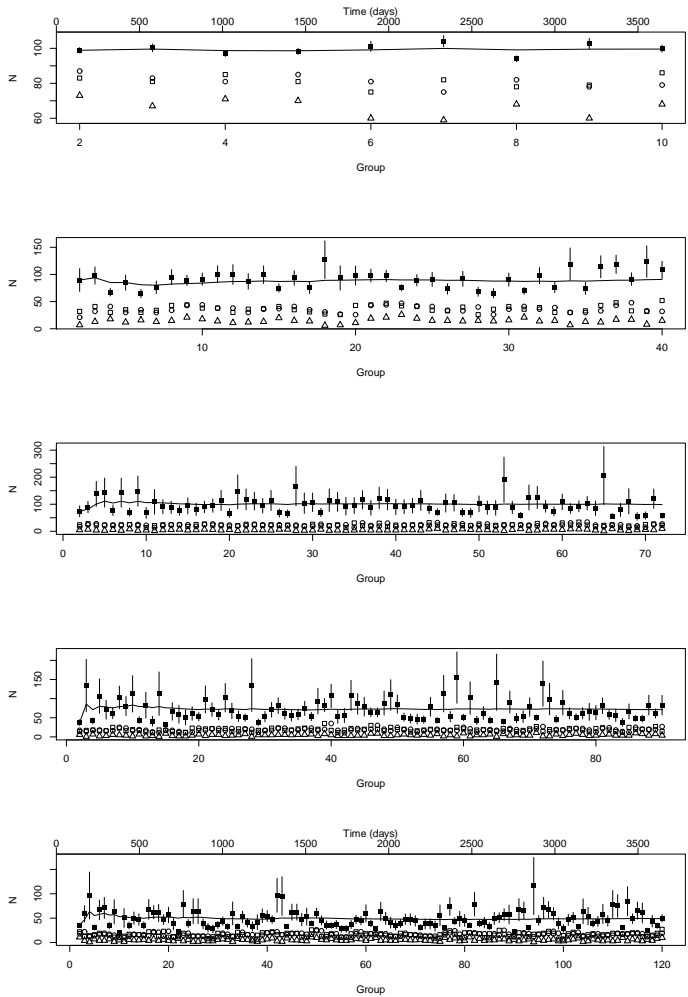}
\caption{Grouped Lincoln-Petersen Estimator applied to simulated decade-long monitoring data for 100 pulsars. The model population has a gaussian \ppuls distribution with $\bar{P}$=200s, $\sigma_{P}$=50s, and is sampled at 7-14 day intervals, with {\it thresh}=0.5. The five panels show results of grouping the observations. From top down: 10 groups of 36, 40 groups of 9, 72 groups of 5, 120 groups of 3. The MLPE value and 1$\sigma$ uncertainty are plotted by solid points, number of pulsars detected in the sample ($N_2$) by open circles, number of pulsars in the previous sample ($N_1$) by open squares, and the number of pulsars in both samples ($N_{12}$) by open triangles. The cumulative running mean is shown as a black line. }
\label{fig:grouped_lpe}
\end{center}
\end{figure}

An improvement is to merge consecutive samples in order to meet the requirements of independence and equal capture probability. This is done by grouping observations and then treating each group as an independent sample. Each group must span a time interval over which one is able (in principle) to capture any individual in the population. This implementation does a much better job, as shown in Figure~\ref{fig:grouped_lpe}, and is highly applicable in the case that one can construct only 2 such groups; to compare two observing seasons for example.

This most basic capture-recapture methodology, in addition to sub-optimal treatment of multiple samples, does not exploit the unique identification of individuals possible in the case of most astronomical objects. This is understandable, as small animals, birds and fish (historically the populations studied) of the same species all look alike. 

\subsection{Cumulative Number Count versus the Schnabel Estimator}
\label{sect:cumulative}
The total number of pulsars or other transient sources detected in an ongoing monitoring campaign naturally rises with time and number of observations. The actual rate of new discoveries is determined by a complex interplay of observing  cadence and population.  Use of expensive telescope time could be optimized by monitoring strategies that are able to determine the total population in the minimum amount of time. 

For example if the first observation in a series detects pulsars ``A, B, and C", the second detects pulsars ``A, C, D", and the third ``B, D, E": then the cumulative total is 4 pulsars.  An extremely interesting question to ask, is how many observations are needed before we have seen every pulsar at least once. Going further, can Capture-Recapture methods accelerate the process? This has an obvious implication for the design of monitoring programs if we can predict the optimum observing cadence and duration of the campaign: If the scientific goal is to find $N$, it may be feasible to obtain a robust estimate without actually detecting all $N$ objects.

A Monte Carlo approach can provide confidence limits on the detection rate for a variety of target model populations. At each iteration, the code generates a fresh pulsar population ($N$ objects) and sampling pattern $S$. It then accumulates the total number of unique pulsars detected as a function of time, and records the observation number $S_{i}$ at which the count exceeded {\it i}\% of $N$ . After 1000 iterations, the results are sorted to produce the probability distribution of $S_{i}$, from which confidence intervals may be drawn. For example: ``For population model X,  sampling cadence $S$ will have detected 75\% of the population after Z observations, with 95\% confidence".

The Schnabel estimate $N_{Sch}$ (see Equation~\ref{eqn:schnabel} is derived from the cumulative running average of new to previously encountered individuals. It is therefore expected to perform better than $N_{LP}$, and to provide an accelerated prediction of the cumulative number count. 
 
\begin{figure*}
\begin{center}
\includegraphics[width=7.5cm, angle=-90]{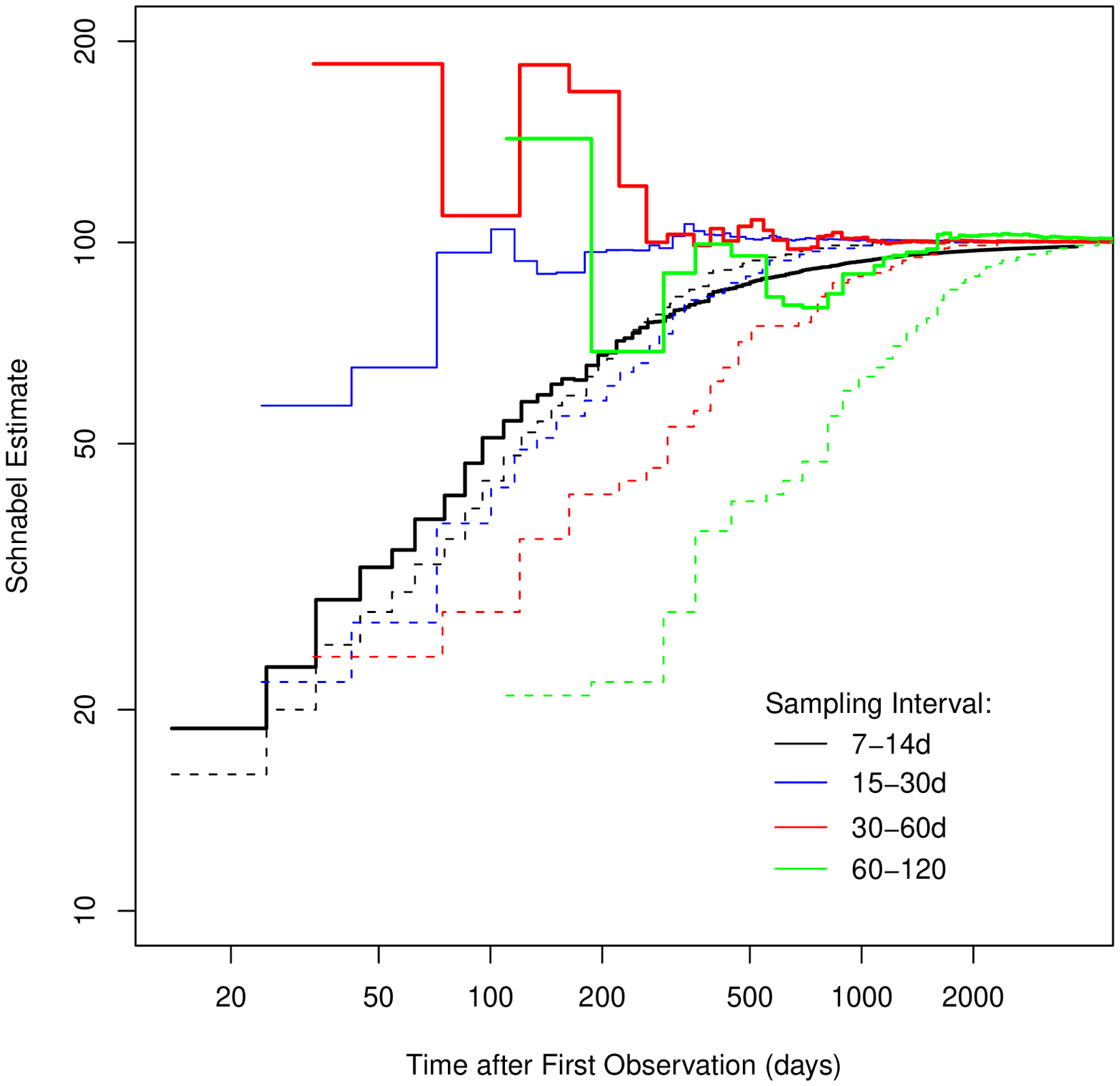}
\includegraphics[width=7.5cm, angle=-90]{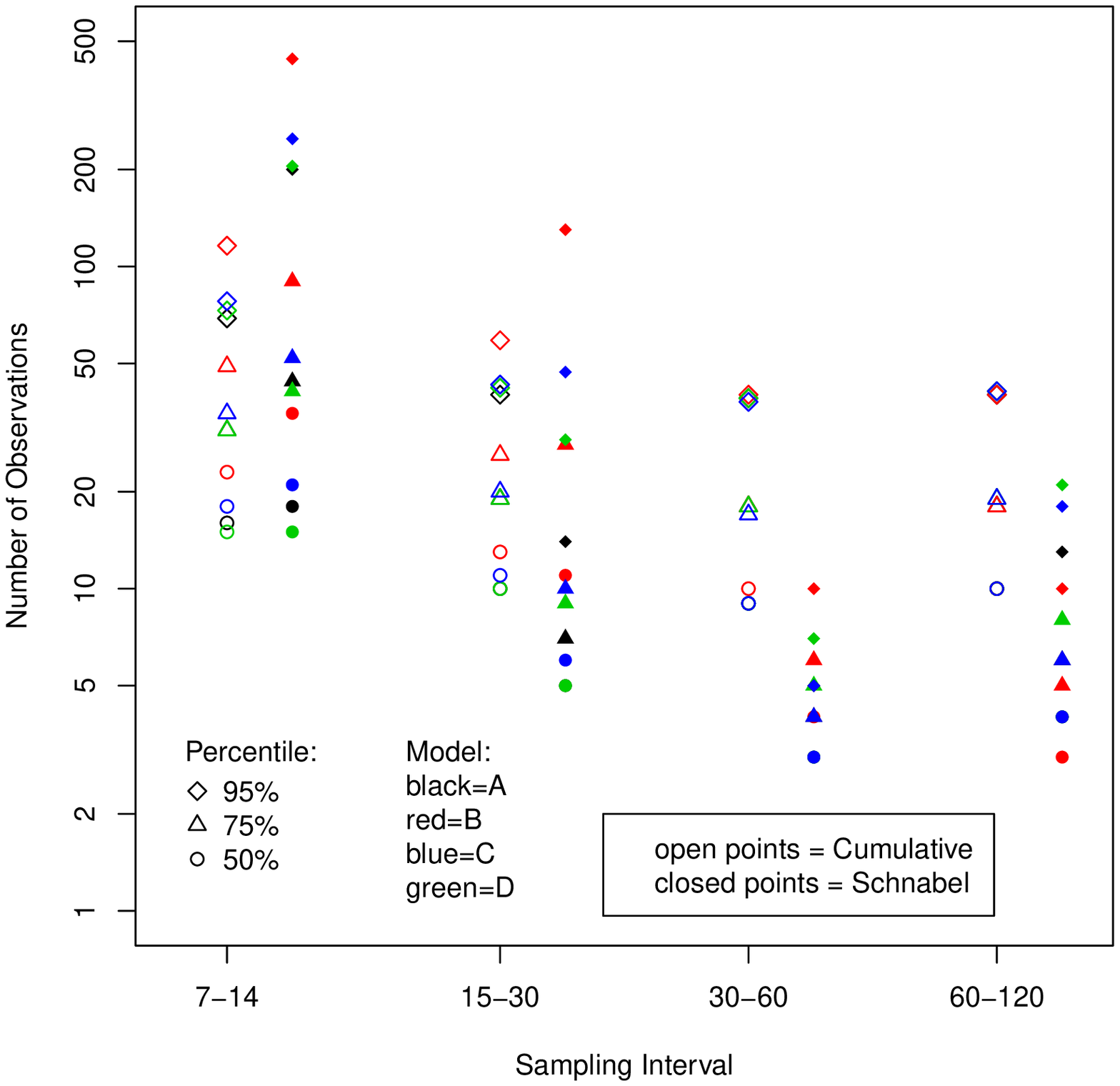}
\caption{Schnabel Estimator compared with cumulative number of pulsars detected as a function of time, for a variety of sampling cadences and population models. {\em Left}: Schnabel estimate (solid) and cumulative number count (dashed) for a gaussian input pulse period distribution with N=100, $\bar{P}$=150s, $\sigma$=50s.  {\em Right}: The number of observations required to recover different target fractions (50\%, 75\%, 95\%) of the population, with 95\% confidence for 4 contrasting population models. The Schnabel estimates generally converge on the true population value in less than half the time taken to discover all of the sources. }
\label{fig:schnabel}
\end{center}
\end{figure*}

We ran a large grid of simulations for $N_{Sch}$ to test a range of observing strategies and population models.
The sampling patterns were 10 year-long stretches of 7-14, 15-30, 30-60, 60-120 day intervals. The population models are given in Table~\ref{tab:models}. Each combination of sampling and model was run 1000 times, and the results summarized in Figure~\ref{fig:schnabel}.  When the sampling density is high compared to the average pulsar duty cycle, $N_{Sch}$ basically track the cumulative count. This follows from the fact that closely spaced consecutive samples have a high degree of correlation. As for $N_{LP}$ the observations would need to be grouped into super-samples, such that the requirement of equal capture-probability is restored.   Steadily increasing the observation spacing causes the cumulative count to lag as successively more outbursts go undetected. This is the regime where Equation~\ref{eqn:schnabel} becomes useful, as demonstrated in Figure~\ref{fig:schnabel}.  A set of example sampling runs for model A shows that $N_{Sch}$ converges rapidly on the underlying value of $N$, cutting the elapsed time by up to a factor of 10. A full Monte Carlo simulation was then used to compute the number of observations required to reach 50\%, 75\% and 95\% of the underlying population in 95\% of trials, for each model/sampling combination.  

\subsection{Exponential Model, and recovery of Capture Probability or Duty Cycle}

\begin{figure*}
\begin{center}
\includegraphics[width=7.5cm, angle=-90]{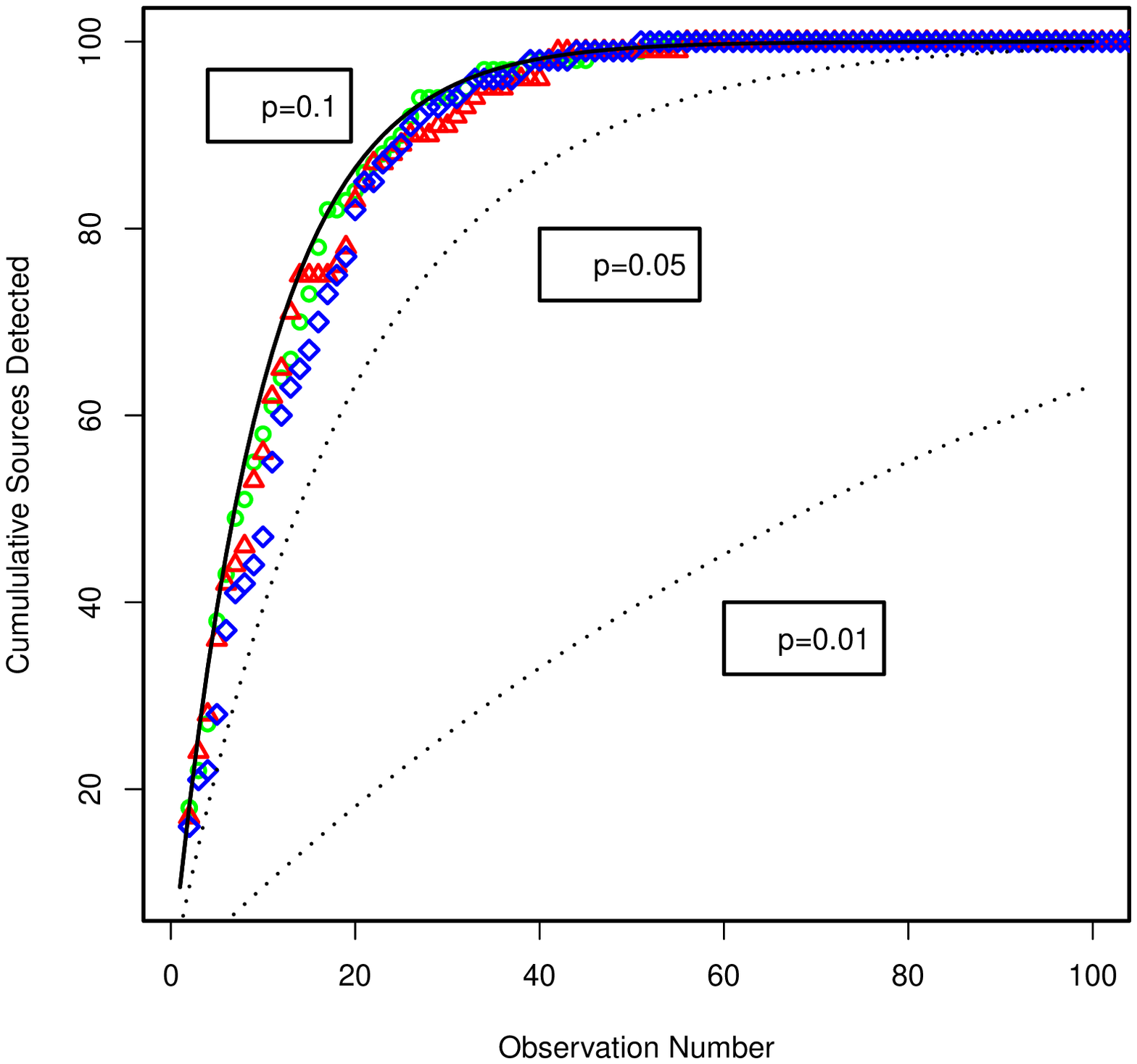}
\includegraphics[width=7.5cm, angle=-90]{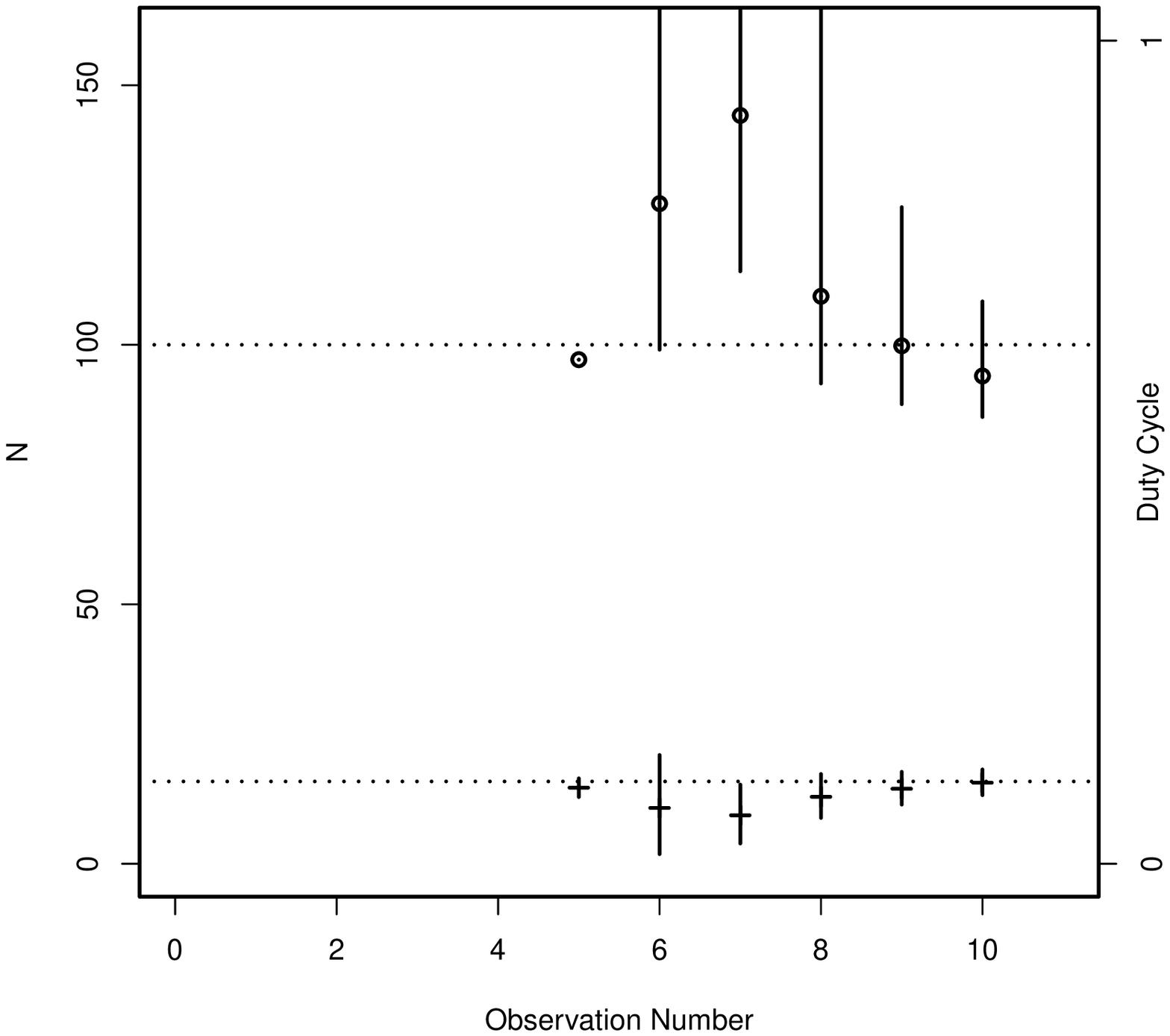}
\caption{The exponential model for capture recapture. Left: Plot shows cumulative total of transients vs observation number, for transient population model A (duty-cycle=10\%) under different sampling cadences (15-30 days, 30-60 days, 60-120 days). The curves are Equation~\ref{eqn:exponential} for $N_0$=100, and various values of $p$. 
Right: Fits to the data converge rapidly on the correct model values of $N_0$=100, $p$=0.1 for a representative set of simulated observations. (Fitting code does not always converge for fewer than 5 points.) }
\label{fig:exponential}
\end{center}
\end{figure*}

As described in section~\ref{sect:equations} a simple equal-probability model for capture-recapture can resemble conceptually the radioactive decay law. The number of un-encountered units in the population falls exponentially with repeated observations, so the model fitting amounts to simultaneously finding $N$ and $p$.  The dual outcomes of the technique are population abundance ($N_0$) {\it and} the encounter probability $p$, which is closely related to the duty cycle. This claim is demonstrated in Figure~\ref{fig:exponential}a, which shows the cumulative number of transients (model A: duty cycle=0.1) under different observing cadences. Interestingly the three cadences (15-30d, 30-60d, 60-120d) all conform to the predicted exponential curve when cumulative source count is plotted against observation number. Equation~\ref{eqn:exponential} is plotted illustrating various values of $p$=0.1,0.05,0.01), and the data-points clearly track the $p$=0.1 curve.  So, the capture history is shown to be independent of time, provided the observations are un-corelated.

Results of fitting Equation~\ref{eqn:exponential} to a simulated capture history are shown in Figure~\ref{fig:exponential}b. This was done in {\em R} using the non-linear least squares function {\it nls} to simultaneously fit for $N_0$ and $p$. Confidence intervals on the fitted parameters were generated by the function {\it confint}. The capture history was fitted for successively increasing number of observations, and the resulting $N_0$ and $p$ (and their 1$\sigma$ uncertainties) plotted against observation number. Fitting for a curve with 2 unknowns requires $>3$ points, and in practice the code needs $>$5, to reliably converge and evaluate confidence intervals. Figure~\ref{fig:exponential}b shows that the exponential model converges on the correct values ($N_0$=100, $p$=0.1) with a 10\% error after $\sim$10 observations.

\subsection{Performance of Advanced Capture-Recapture Algorithms}

The most significant development in Capture-Recapture methods is their extension beyond the restrictions of equal capture probability and closed populations. Capture probability is the likelihood that a given individual will be encountered in any arbitrary sample of the population. Populations are open or closed based on whether individuals can enter and leave. In the life sciences, heterogeneity in capture probability arises from both individual behavior and experimental design, and populations feature births, deaths and migration.   In astronomy we can comfortably ignore migration, and birth/death mostly occurs on timescales much longer than the typical monitoring program, the longest of which exceed the human lifetime only by small factors, so astronomers need only consider methods for closed populations. In terms of capture probability however, sources can show a broad distribution of duty cycle, and not all observations are equally sensitive.  Transient sources are frequently recurrent, or even periodic (e.g. X-ray binaries) or subject to other strong temporal correlations. In the case of novae for example the occurrence of a thermonuclear runaway exhausts the fuel supply, so the probability of recurrence drops to zero for a time. 

In preceding sections, the classical Lincoln-Petersen and Schnabel formulae were shown to work well for astronomical source populations  with fixed $p$. The caveat being that sampling cadence is selected appropriately, or observations grouped prior to analysis.
We now move on to consider closed populations conforming to the exponential or log-linear model \cite{}, and its extension to heterogeneous capture probabilities \citep{darroch1993}. In this context, variation in duty-cycle among individual transients. The goals are to see if robust population estimates are possible, and whether advances in dealing with heterogeneity result in reduced bias and faster (in terms of observing time) convergence. One of the principal software packages for capture-recapture analysis is {\it Rcapture} \cite{rcapture} which handles both open and closed populations. {\it Rcapture} implements classes of models in which the capture probabilities of individuals are allowed to vary during the experiment. Within a closed population, the probabilities can vary with time (t), heterogeneity between individuals (h), and changes in behavior caused by response to capture (b) (\cite{otis1978}). Case b initially seems irrelevant in astronomy, however the `behavioral response'  model is actually an attempt to correct for correlations between neighboring observations. This is analogous to the case of transient source outbursts that last for longer then the spacing between observations.  We tested the {\it Rcapture} model classes $M_t$, $M_h$, $M_b$, $M_{th}$ (heterogeneity and temporal variation) and $M_0$ (no variation). The models are fully described by \cite{rcapture} and were fit by poisson regression to the simulated capture history data. 

Analysis with {\it Rcapture} proceeds by fitting the  plausible models, selecting the one that best represents the data, and using it to estimate abundance and confidence interval.

Capture histories were generated for the usual set of models (Table~\ref{tab:models} and observing strategies, and analyzed with the {\it closedp} and {\it closedp.0} algorithms. {\it closedp} was used to perform poisson regression fitting of heterogeneous and time-dependent models. Due to the computational load, {\it closedp} is currently limited to analyzing $\leq$20 observations at a time. For longer datasets we used {\it closedp.0}  which fits the heterogeneous capture-probability models in non time-dependent form. 

\begin{figure}
\begin{center}
\includegraphics[width=8cm, angle=-90]{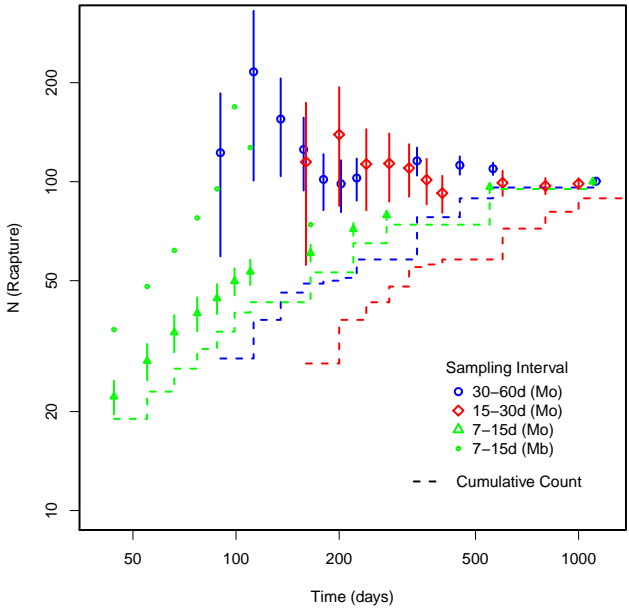}
\caption{  {\it Rcapture} population abundance estimates. The population model is {\it A} (See Table~\ref{tab:models}. Results for three different sampling cadences are plotted. Points with error-bars indicate the $M_0$ fitting results with 1$\sigma$ uncertainties. Dashed histograms show the cumulative number of objects detected in each series.  Result for $M_b$ fitting is plotted for the over-sampled series, showing some compensation for the correlation bias. }
\label{fig:rcapture}
\end{center}
\end{figure}

In the cases where the Schnabel method worked well, for example model A (Table~\ref{tab:models}), with 15-30 day or greater sampling intervals, {\it Rcapture} showed no need to invoke heterogeneity or temporal variation. We deduce this from the fact that $M(0,t,h,th)$ all converged on the correct result within 5-10 observations, as shown in Figure~\ref{fig:rcapture}. In addition the uncertainty estimates become very constraining after only 5 observations.  Shortening the sampling interval into the regime where strong correlation is present between observations (e.g. Model A and 7-15 day intervals) causes bias in {\it Rcapture}'s estimates. As in the Schnabel case the results track the cumulative discovery rate, with very little difference between $M_{0,h,t,th}$. 
The type of serial correlation present in the simulated lightcurves could mimic the behavioral response model $M_b$. A star detected in one observation is likely to be seen in the following observation(s) if the sampling interval is sufficiently short. This was tested for models A and C (a population dominated by short recurrence periods), and indeed $M_b$ abundance estimates do a much better job in the dense sampling regime (in the sparse sampling regime $M_b$ did not differ strongly from $M_0$).

Next, variation in source behavior was introduced by population simulations having a broader distribution of recurrence periods. Model D (Table~\ref{tab:models}) has double the width compared to model A, and Model B is flat in \ppuls, leading to a monotonically rising distribution of \porb (i.e. recurrence time).
{\it Rcapture} exhibited only small (i.e. smaller than the errors) differences in abundance estimators for $M_{h}$ and the uniform capture-probability c$M_0$, for models B, D compared to model A. This result demonstrates that capture-recapture provides robust population size estimates in the face of types of population diversity expected in astronomy. 

Efficient use of densely (over)sampled monitoring data is achieved by grouping or merging the observations (as was pointed out in Section~\ref{sect:lpe}).  In {\it Rcapture} large datasets can be merged in this way using the function {\it periodhist}. 

\section{Conclusions and Further Applications}

Simple formulae for Capture-Recapture provide meaningful results provided the observing strategy is matched to the characteristic recurrence timescale of the target population. Modern methods provide robust estimates for all populations and observing cadences investigated in this study. Our generalized exponential population model is simple to implement, performs better than the classical methods, and is a bridge to the more sophisticated methods developed in the life-sciences. 

The modified Lincoln-Petersen index is useful when only 2-3 observations are available. It can be extended to longer series by constructing a running mean, or by merging groups of contiguous observations. The latter strategy is valuable in the cases of seasonable observability, and observations with very low detection rates. This class of estimators were found to be strongly biassed by sampling frequency and tend to converge on an underestimate of the population. 

The Schnabel estimator is close to ideal for simple analysis and modeling. It converges on the true population rapidly, enabling the population size to be estimated in a fraction of the time required to discover all individuals. This property is extremely useful for predicting the performance of proposed monitoring strategies, and tracking progress in ongoing studies. Figure~\ref{fig:schnabel} shows that the Schnabel estimator can provide a dramatic acceleration in population estimation, potentially reducing both on-source telescope time and monitoring duration by 50-90\%. Confidence interval estimation is readily amenable to monte-carlo methods.

Capture-Recapture computational algorithms developed for the life sciences provide a powerful set of generalized poisson regression models to fit a wide variety of monitoring data. The package {\it Rcapture} implements the most applicable modern methods in $R$ making it easily accessible to users unfamiliar with the ecology and medical settings where most of the methods were developed. The log-linear modeling framework implemented in {\it Rcapture} has three advantages: (1) Extends the capture-recapture paradigm beyond its classical domain of validity. (2) Provides robust confidence intervals (3) Use of generalized models avoids time-consuming generation of problem-specific models.  We did not investigate the robust design methodology introduced in {\it Rcapture}, which combines periods of study during which the population is treated as closed, separated by unobserved periods, during which it is open. Qualitatively such an approach could well apply to astronomical datasets, where seasonal observability is a factor, or where short-term activity is superimposed on slower trends. For example periodic X-ray binary outbursts fueled by a circumstellar disk that grows and dissipates on a scale of years. We note that \cite{Romine2016} have recently used {\it Rcapture} to study the association of ultra hard X-ray sources with protostars, an application that does not involve the time domain at all. 

Computational methods were found to outperform simple estimators as soon as the number of observations exceeds 3-5. For all astronomical source populations simulated in this study {\it Rcapture} converged on the correct abundance (estimate and errors within 10\% of the true value) within 8-15 observations.  Depending only slightly on the sampling cadence and model, this generally translated to a point at which about half the population had been encountered. The principal objection to applying techniques from other disciplines (in this case biology) is that methods may be used outside their domain of applicability, leading to unreliable results.  Such concerns were thoroughly investigated in this paper using realistic numerical simulations of astronomical monitoring campaigns.  

The ideal astronomical setting for Capture-Recapture analysis is a fixed population observed on multiple occasions, separated by sufficiently long intervals for different subsets to be active. In addition to estimating population abundance, the approach can provide frequency or duty cycle for rare states in continuously visible sources. In addition to the example of flare stars \citep{Ambartsumian1975}, the frequency of dust-shell ejections in R Corona Borealis stars and horizontal branch giants, and outbursts of the Be phenomenon are all good candidates.   

We studied periodic transient sources in this paper, precisely because they appear (at first glance) to diverge strongly from the first axiom of the Capture-Recapture paradigm: equal capture probability. Individuals whose activity patterns are primarily periodic do not have a purely random chance of being seen in any arbitrary sample. Their activity is correlated. Nevertheless in Section~\ref{sect:results} we found that for assemblies of randomly distributed periodic sources, Capture-Recapture estimators remain valid if sampling cadence is appropriate. Aperiodic transients on the other hand are expected to recur at random, and to exhibit almost perfectly uniform (with time) capture probabilities. We say {\it almost}, because outburst duration, brightness (i.e. heterogeneity in peak luminosity and/or distance), and typical recurrence interval all affect detection probability.  Black hole X-ray novae (accreting binaries containing a black hole and a low mass star) have a very poorly known space density due to their rarity. Several concerted efforts to determine the space-density and duty cycles of black-hole transients are underway, e.g. ChaMPlane (in the Milky Way) and are a motivating factor for the approach investigated here.

Capture-Recapture methods have the potential for wide application in time domain astronomy, and efforts to apply the methods explored here are underway. A number of monitoring studies of X-ray transients in external galaxies are currently planned, in progress, or recently completed (see introduction). At optical wavelengths large numbers of recurrent transients are seen by time-domain surveys such as OGLE (Magellanic Clouds) and \cite{ogle}, MACHO (the Galactic Bulge) \cite{macho}. New all-sky surveys include PanStars \cite{panstars}, Palomar Transient Factory \cite{ptf} and LSST \cite{lsst}. The photographic plate digitization project DASCH will soon access $\sim$100 years of well sampled photometric data for the entire celestial sphere.  New statistical tools are needed, to tackle questions that could not be asked prior to this explosion of transient sources.

\section*{Acknowledgements}
SL thanks the anonymous referee for constructive suggestions on python compatibility, and for bringing to our attention the work of V. A. Ambartsumian. This work was supported in part by NASA Astrophysics Data Analysis Program grant NNX14-AF77G.

\section{Appendix: Calling R functions from within Python}
The R programming language is the standard for statistical analysis and data visualization/exploration in the life sciences. In their book {\it Modern Statistical Methods for Astronomy}, \cite{FB} provide many example applications of available {\it R} functions to astronomical research. R is a very high level (i.e. easy to use), object oriented language with an intuitive and streamlined syntax compared to other languages (e.g. C, Fortran, IDL etc). Pre-defined functions are contributed by an active user community covering the gamut of statistical tests, time domain (e.g. FFT), kernel smoothers, bootstrap estimation, and interactive 3D visualization.  

Users of {\it Python} can call {\it R} functions directly using the {\it rpy2} function which creates an instance of R that runs within python, so with a few simple lines of code, all of the functionality of {\it R} analysis and plotting can be done within a {\it Python} framework.\\

\noindent {\tt > pip install rpy2}   -  puts the {\it rpy2} package on the machine. \\
{\tt > import rpy2 }   -  this imports the entire package into the program \\  
{\tt > import rpy2.robjects as robjects}  - Import the module {\it robjects} which calls the R code within the Python program \\
{\tt > R = robjects.r }  -  this creates an alias R for robjects.r   \\
{\tt > result = R.[function](?) }  \\
{\tt > from rpy2.robjects.packages import importr} - this line gives you the ability to import r packages\\
{\tt > capture = importr("Rcapture") } - this imports the Rcapture package and assigns the alias capture to it\\

To call the routine {\it closedp} from {\it Rcapture}: \\

{\tt > capture.closedp(`YourData')}   where ``YourData" is a capture history: an array of 1s and 0s wherein each row represents an observation (or occasion) and each column represents a unique individual. 1 indicating the individual is present on that occasion, otherwise zero.\\

\noindent  Here is a comparison of the performance of Equations 7 and 8. Using  an example capture history (format is a cumulative count of new transients in a series of observations) generated by the {\it Blackbirds} code.

\noindent{\tt > x $<-$ c(1:10); y $<-$ c(17,24,28,36,42,44,46,53,56,62) }\\
{\tt > model $<-$ nls(y $\sim$ a*(1-exp(-b*x)), start = list(a=10, b=0.5)) } - fit Equation 7 \\
{\tt > model2 $<$- nls(y $\sim$ a*(1-  (1 - b)\^x ), start = list(a=10, b=0.5)) } - fit Equation 8 \\
{\tt > coef(model) }  - obtain the values of the coefficients from the object containing the result of the fit \\ 
{\tt > confint(model,levels=0.1,0.9) } - obtain the 95\% confidence intervals from the object containing the result of the fit.


\label{lastpage}

\end{document}